\documentclass{arx}
 
\JournalSubmission   % uncomment for Conference submission

\usepackage{microtype}                 % use micro-typography (slightly more compact, better to read)
\PassOptionsToPackage{warn}{textcomp}  % to address font issues with \textrightarrow
\usepackage{textcomp}                  % use better special symbols
\usepackage{mathptmx}                  % use matching math font
\usepackage{times}                     % we use Times as the main font
\usepackage{tabu}                      % only used for the table example
\usepackage{booktabs}
\usepackage{enumerate}
\usepackage{enumitem}
\usepackage{float}
\usepackage{dirtytalk}
\usepackage{csquotes}
\usepackage{amsmath}
\usepackage{amssymb}
\usepackage{subcaption}
\usepackage{bbm}
\usepackage[export]{adjustbox}
\usepackage{todonotes}
\usepackage{siunitx}
\usepackage{mathtools}
\usepackage{url}
\usepackage{algorithm}
\usepackage[noend]{algpseudocode}
\usepackage{dutchcal} 
\usepackage{multirow}

\DeclareMathAlphabet{\mathcal}{OMS}{cmsy}{m}{n}
\newcommand{\mat}[1]{\mathbf{#1}}

\newcommand{\func}[1]{\mathrm{#1}}

\newcommand{\N}{\mbox{\rm \hbox{I\kern-.15em\hbox{N}}}}

\newcommand{\of}[1]{\!\left( #1 \right)}

\newcommand{\norm}[1]{\left\Vert {#1} \right\Vert}

\newcommand{\Sec}[1]{Section~\ref{sec:#1}}
\newcommand{\App}[1]{Appendix~\ref{sec:#1}}
\newcommand{\Fig}[1]{Figure~\ref{fig:#1}}
\newcommand{\Tab}[1]{Table~\ref{tab:#1}}

\newcommand{\Eq}[1]{Equation~\eqref{eq:#1}}
\newcommand{\Alg}[1]{Algorithm~\ref{alg:#1}}

\algrenewcommand\textproc{}
\algnewcommand{\LineComment}[1]{\vspace{1mm}\State \textit{\small // #1}}

\usepackage[T1]{fontenc}
\usepackage{dfadobe}

\usepackage{cite}  % comment out for biblatex with backend=biber
\BibtexOrBiblatex
\electronicVersion
\PrintedOrElectronic
\usepackage{graphicx}

\title[NePHIM]{\hspace{1.7cm} NePHIM: A Neural Physics-Based \newline Head-Hand Interaction Model}

\author[Wagner et al.]{
{\parbox{\textwidth}{\centering Nicolas Wagner$^{1,2}$, Ulrich Schwanecke$^2$, and Mario Botsch$^1$
         }       
}
\\
{\parbox{\textwidth}{\centering $^1$TU Dortmund University, Germany \\
                                $^2$University of Applied Sciences RheinMain, Germany    }
}
}
\begin{document}

\teaser{
    \centering
    \includegraphics[width=\linewidth]{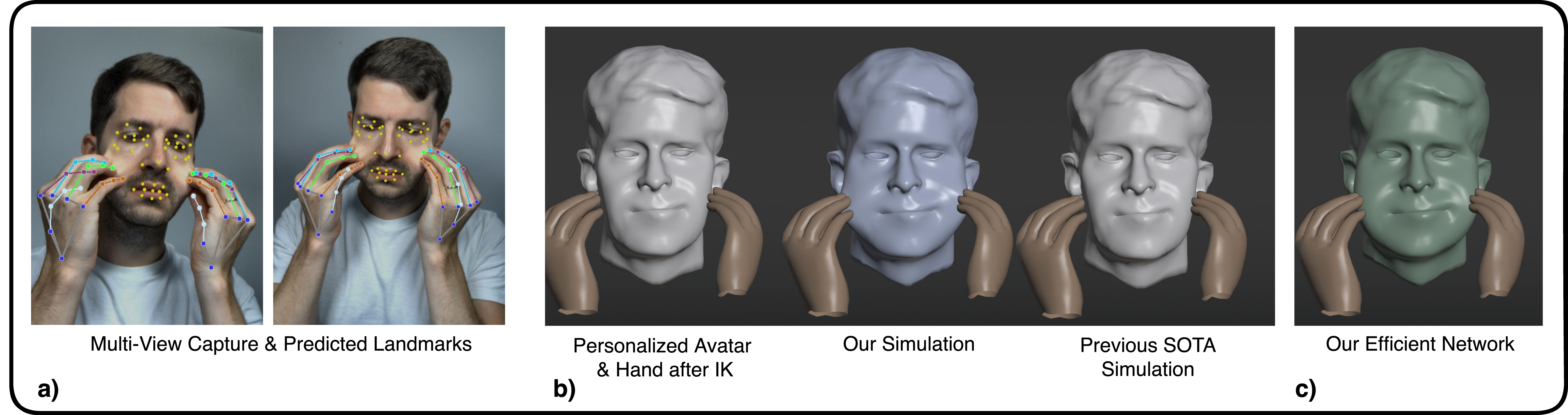}
    \caption{The different steps of NePHIM by means of a single frame: a) Two of  the 16 views of our multi-camera rig used to capture head-hand interactions and corresponding landmarks. b) Our proposed simulation for head-hand interactions in comparison to the tracked input (after fitting template surfaces to the landmarks (IK)) and the simulation used in the previous state-of-the-art \cite{shimada2023decaf}. c) Prediction of the efficient neural network trained to approximate our simulation.}%
    \label{fig:teaser}%
    \vspace{0.51cm}
}

\maketitle
%-------------------------------------------------------------------------
\begin{abstract}
Due to the increasing use of virtual avatars, the animation of head-hand interactions has recently gained attention. To this end, we present a novel volumetric and physics-based interaction simulation. In contrast to previous work, our simulation incorporates temporal effects such as collision paths, respects anatomical constraints, and can detect and simulate skin pulling. As a result, we can achieve more natural-looking interaction animations and take a step towards greater realism. However, like most complex and computationally expensive simulations, ours is not real-time capable even on high-end machines.  Therefore, we train small and efficient neural networks as accurate approximations that achieve about 200 FPS on consumer GPUs, about 50 FPS on CPUs, and are learned in less than four hours for one person. In general, our focus is not to generalize the approximation networks to low-resolution head models but to adapt them to more detailed personalized avatars. Nevertheless, we show that these networks can learn to approximate our head-hand interaction model for multiple identities while maintaining computational efficiency.

Since the quality of the simulations can only be judged subjectively, we conducted a comprehensive user study which confirms the improved realism of our approach.
In addition, we provide extensive visual results and inspect the neural approximations quantitatively.  All data used in this work has been recorded with a multi--view camera rig and will be made available upon publication. We will also publish relevant implementations. 
%-------------------------------------------------------------------------
%  ACM CCS 1998
%  (see https://www.acm.org/publications/computing-classification-system/1998)
% \begin{classification} % according to https://www.acm.org/publications/computing-classification-system/1998
% \CCScat{Computer Graphics}{I.3.3}{Picture/Image Generation}{Line and curve generation}
% \end{classification}
%-------------------------------------------------------------------------
%  ACM CCS 2012

%The tool at \url{http://dl.acm.org/ccs.cfm} can be used to generate
% CCS codes.
%Example:


%ARXIV \printccsdesc   
\end{abstract}  
%-------------------------------------------------------------------------
%-------------------------------------------------------------------------

\section{Introduction}

\label{sec:int}
How many times per hour do you think you touch your face? Probably more often than you are aware of. Although the answer to this question varies in scientific studies, it can be said that, on average, people touch their heads several dozen times an hour~\cite{kwok2015face, rahman2020frequently, mueller2019self}. There are many ways to interact, such as touching, stroking, scratching, rubbing, pulling, tugging, squeezing, and caressing, to name but a few. Behavioral sciences do not conclusively answer why people touch their faces, yet the implications even extend to computer graphics. Due to the frequency and expressiveness of head-hand interactions, simulating them in facial animations would considerably improve user perception. Especially with the focus on photo-realistic avatars these days~\cite{qian2024gaussianavatars, ma2024gaussianblendshapes, zielonka2023instant, athar2022rignerf, garbin2022voltemorph}, the relevance of authentic facial animations is further accentuated.

Only recently, attempts to incorporate head-hand interactions into facial animations have been proposed~\cite{shimada2023decaf, wu2024dice}. In particular, these approaches address two main challenges. Naturally, the simulation of interactions is the main emphasis, but three-dimensional tracking of the head and hands is also a prerequisite for realistic animations. Shimada et al.~\shortcite{shimada2023decaf} and Wu et al.~\shortcite{wu2024dice} are impressive in determining simulated 3D head and hand surfaces from a single monocular image. However, both neglect the fidelity of the interaction animation as they are based on the same rather coarse physics-based \emph{surface} simulation. More sophisticated, detailed, and anatomically accurate \emph{volumetric} physics-based simulations of heads have been explored in other contexts \cite{sifakis2005automatic, ichim2017phace, cong2016art, chandran2024anatomically}.

This work introduces a substantially improved physics-based simulation of head-hand interactions and designs more realistic interaction mechanisms. For instance, in contrast to the previous methods, we consider pulling interactions and the influence of the skull. Since this simulation is not real-time capable, we also learn a personalized neural network as an approximation. Both our simulation and the network process tracked head and hand surfaces and thus remain compatible with the tracking concepts of previous approaches~\cite{shimada2023decaf, wu2024dice}. Another contribution of this work is the creation of a dataset of real head-hand interactions. To this end, we built a multi-view rig with 16 high-resolution and synchronized video cameras with which we recorded several subjects. Unlike the only other comparable dataset available~\cite{shimada2023decaf}, we did not instruct the participants which head-hand interactions to perform. We simply asked the participants to perform arbitrary interactions and can, therefore, reproduce an even wider range of hand movements in our data. Among other things, we also capture skin pulling, which was previously ignored. \Fig{teaser} is an exemplary illustration that shows a recorded \textit{pulling} frame, the associated simulation, and the approximation by our neural network.

We evaluate our approach qualitatively using visual examples and the accompanying video of dynamic head-hand interaction animations. Furthermore, we conducted an extensive user study that confirms that our approach is perceived more naturally than previous ones. Quantitative experiments demonstrate that the neural approximation can be created in just a few hours and adapted to multiple human identities simultaneously. The trained network achieves around 50 frames-per-second (FPS) even on slower CPUs.
\section{Related Work}\label{sec:rw}
In this section, we discuss three literature fields related to our approach. First, \Sec{rw:py} presents physics-based facial animations in general. Next, \Sec{rw:hh} addresses recent developments focusing specifically on animated head-hand interactions. Finally, \Sec{rw:lpy} examines work in which neural networks approximate physics-based simulations.

\subsection{Physics-Based Facial Animations}\label{sec:rw:py}
Heuristic physics-based facial simulations have been developed for a long time and principally intend to compensate for shortcomings of \emph{simpler} but popular facial animation methods like linear blendshapes~\cite{lewis2014practice}. 
For instance, artifacts like implausible contortions and self-intersections can be avoided by including volumetric and anatomical constraints. The pioneering work of Sifakis et al.~\shortcite{sifakis2005automatic} is a volumetric phy\-sics-based facial simulation that runs on a personalized tetrahedron mesh. Unfortunately, the tetrahedral mesh can only be of limited resolution due to an associated dense optimization problem. With Phace \cite{ichim2017phace, ichim2016building}, an improved simulation concept has been introduced, which is also defined on a tetrahedral mesh but can handle higher resolutions and considers anatomical structures more precisely. In addition to a tetrahedral mesh, the art-directed muscle models \cite{cong2019muscle, bao2019high, cong2016art} represent muscles as B-splines that steer facial expressions via trajectories of spline control points. A solely inverse model for determining the physical properties of faces was proposed in \cite{kadlevcek2019building}. 

Thanks to increased computing capabilities, data-driven physics-based facial simulations have also become appealing recently. An example is the model of Yang et al.~\cite{yang2022implicit} that learns to volumetrically animate a person's face from multi-view videos with the help of differentiable physics \cite{du2021diffpd}. Although Yang et al.~\shortcite{yang2023implicit} extend the model to cover several identities, adding a novel identity requires five days of retraining and the inference of one frame runs multiple seconds. Generally, heuristic as well as data-driven physics-based simulations are not commonly used in real applications due to their complexity and computational effort. Other data-driven approaches include Animatomy~\cite{choi2022animatomy}, which represents muscles as curves, and the implicit model of Chandran et al.~\shortcite{chandran2024anatomically}. The aforementioned data-driven simulations are not designed to handle collisions and external interactions.

\subsection{Head-Hand Interactions}\label{sec:rw:hh}
All previously discussed simulations have in common that they are primarily aimed at facial animation, facial retargeting, or face reconstruction, but not at the simulation of external influences such as hands. Although models like Phace~\cite{ichim2017phace} are theoretically applicable in such scenarios, the non-trivial practical implementation of interactions has not happened until lately. Shimada et al.~\shortcite{shimada2023decaf} propose the first head-hand interaction simulation, Decaf, and demonstrate how a neural network can learn the simulation while generalizing over the FLAME head model~\cite{li2017learning} and the MANO hand model~\shortcite{MANO:SIGGRAPHASIA:2017}. Decaf focuses on mapping a single RGB image to interaction deformations, using only a surface-based simulation that, in terms of quality and realism, falls short of the volumetric simulations discussed in \Sec{rw:py}. Also, the low resolution and the sometimes too smooth representation of heads in FLAME are often insufficient for demanding applications. Wu et al.~\shortcite{wu2024dice} advance Decaf by an extended generalization to in-the-wild images, unfortunately, the underlying simulation remains the same. Consequently, we focus on a more realistic simulation for personalized and more detailed head avatars.

\subsection{Approximating Physics-Based Simulations}\label{sec:rw:lpy}
As we accelerate our approach with efficient neural networks, we also give a brief overview of the literature on neural approximations of physics-based simulations. On the one hand, there are general methodologies~\cite{srinivasan2021learning} that also explicitly deal with interactions of two or more objects~\cite{romero2022contact, romero2021learning}. On the other hand, there are methods with a focus on bodies \cite{santesteban2020softsmpl, casas2018learning} or heads~\cite{wagner2023softdeca}. For NePHIM, we adopt the general method of subspace neural physics~\cite{holden2019subspace} that is, in particular, computationally efficient for approximating simulations of interacting objects.
% ===========================================================
% ===========================================================
% Notation
\newcommand{\tm}{t} % Time step
\newcommand{\head}[2]{$H^{#2}_{#1}$} % Head
\newcommand{\lhand}[2]{$L^{#2}_{#1}$} % Left Hand
\newcommand{\rhand}[2]{$R^{#2}_{#1}$} % Right Hand
\newcommand{\track}{\mathrm{tra}} % Tracked
\newcommand{\reg}{} % Tracked
\newcommand{\inter}{\mathrm{int}} % Interactive
\newcommand{\corr}{\mathrm{cor}} % Corrected
\newcommand{\lmk}{\mathrm{lmk}} % Landmarks
\newcommand{\sib}{sb} % Simulated Before
\newcommand{\sia}{\texttt{phy}} % Simulated After
\newcommand{\pre}{\texttt{net}} % Simulated After
\newcommand{\proj}[1]{\tilde{#1}} % Simulated Prediction
\newcommand{\alg}[1]{\texttt{#1}} % Simulated Prediction
\newcommand{\tet}[1]{\mathbb{#1}} % TetMesh
\newcommand{\elm}[1]{\mathsf{#1}} % Element
\newcommand{\sur}[1]{\func{s}\of{#1}} % Vertices
\newcommand{\pca}[1]{\lowercase{#1}} % PCA
\newcommand{\sq}[1]{\mathcal{#1}_T} % Sequence
\newcommand{\tmp}[1]{\mathbf{#1}} % Template
% ===========================================================
% ===========================================================
\section{Method}\label{sec:mtd}
This section first outlines the objectives of our approach (\Sec{mtd:goals}) and then presents the formal implementation (Sections \ref{sec:mtd:temp}--\ref{sec:mtd:net}).
To support the reading flow, we slightly misuse the notation in the following derivations by denoting meshes and the corresponding vector of stacked vertex positions with the same symbol. \Tab{not} gives a summary of the notation.

\begin{table}[]
\centering
\begin{tabular}{ll}
\hline
Variable & Description \\ \hline
    $\tet{S}, \tet{J}, \tet{C}$ & Tetrahedral meshes  of soft tissue, jaw, \\ & and cranium        \\
    $H, L, R, J, C$     & Surface meshes of head, left hand, right hand,  \\ & jaw, and cranium       \\
     $E_{*}$ & Energies         \\
     $w_{*}$ & Scalar weights         \\
    $C_{*}$ & Vertex targets of hand interactions  \\
    $I_{*}$ & Set or dictionary of vertex indices \\
    $*_t$ & Indicates the time step $t$ of a variable  \\

    $*^\mathrm{src}$ & Indicates the source $\mathrm{src}$ of a variable  \\
    $\sq{X}$  & Sequence $\of{X_\tm}_{\tm = 1}^T$ of $T$ surface meshes $X_\tm$ \\
    $\proj{X}$     & Projection into PCA space of surface mesh $X$         \\
    $\elm{v}, \elm{c}, \elm{t}$ & A geom. element like a vertex $\elm{v}$, a cylinder $\elm{c}$,\\ & or a tetrahedron $\elm{t}$ \\
    $\texttt{func}$ & Denotes a function 
    \\ \hline
\end{tabular}
\caption{The notation of the main concepts of \Sec{mtd}.}
\label{tab:not}
\end{table}

\subsection{Objectives \& Method Overview}\label{sec:mtd:goals}
\begin{figure}[tp]
    \centering
    \includegraphics[width=1.0\linewidth]{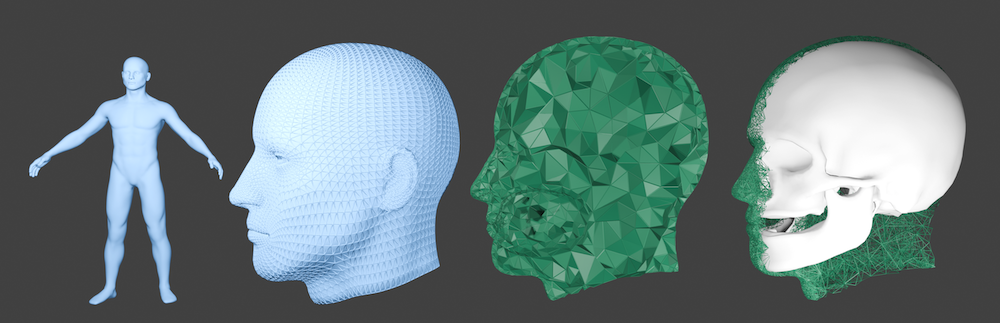}
    \vspace{-0.6cm}\caption*{\raggedright \hspace{1.0cm} a) \hspace{1.4cm} b) \hspace{1.6cm} c) \hspace{1.7cm} d)} \vspace{-0.2cm}
    \caption{a) Full-body template which includes the head template surface $H$ shown in b). c) Cross section of the connected tetrahedral meshes $\tet{S}, \tet{J}, \tet{C}$. d) The template jaw and cranium surfaces $J,C$ embedded in the tetrahedral meshes.}
    \label{fig:temp}
\end{figure}
\begin{figure*}
    \centering
    \includegraphics[width=1.0\linewidth]{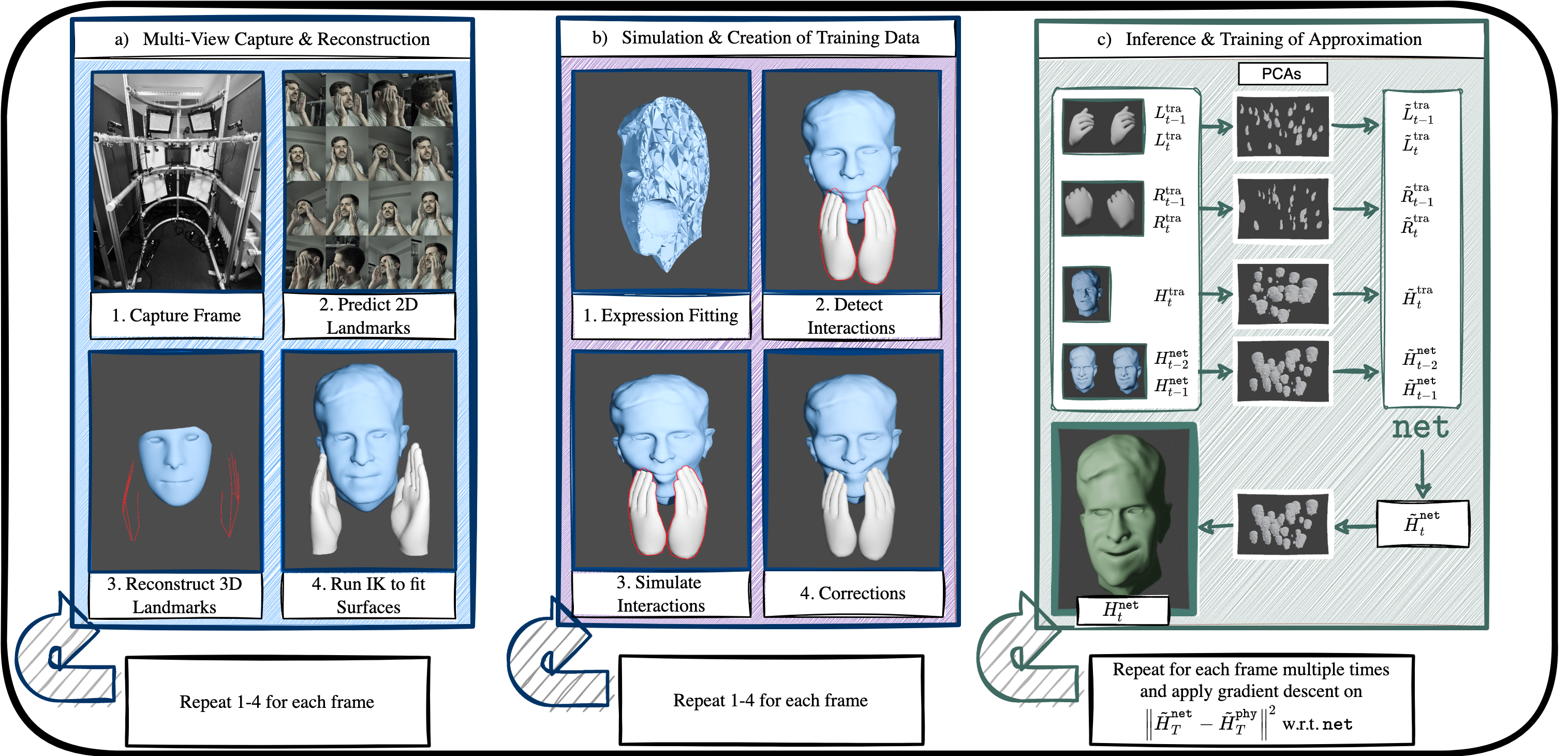}
        \caption{Overview of the three stages of our approach. a) Data capturing as described in \Sec{res:data}. b) All steps of our physics-based simulation $\texttt{phy}$ as explained in \Sec{mtd:sim}. c) Efficient neural approximation \pre~of \sia~as explained in \Sec{mtd:net}.}
    \label{fig:method}
\end{figure*}
We consider an animation at time $T$ with tracked surfaces for the left hand $L^{\track}_{T}$, the right hand $R^{\track}_{T}$, and the head $H^{\track}_{T}$ of a person. Given the corresponding neutral head surface mesh $H$ as well as tracked sequences (consisting of all previous frames up to $T$) for the left hand $\sq{L} = \of{L^{\track}_\tm}_{\tm = 1}^T$, right hand $\sq{R} = \of{R^{\track}_\tm}_{\tm = 1}^T$, and head $\sq{H} = \of{H^{\track}_\tm}_{\tm = 1}^T$, our first objective is to deform the tracked head surface mesh at time $T$, $H^{\track}_{T}$, to
\begin{equation}\label{eq:phy}
\begin{aligned}
    H^{\sia}_{T} = \sia\of{\sq{R}, \sq{L}, \sq{H}, H}, \\
\end{aligned}
\end{equation}
such that head-hand interactions are resolved \emph{realistically} through a physics-based simulation \sia. Previous methods \cite{shimada2023decaf, wu2024dice} determine deformations through a simple surface-based simulation \cite{muller2007position} incorporating only constraints for the skin surface and (static) pushing hand interactions. We improve realism by implementing \sia~(\Sec{mtd:sim}) as a \emph{volumetric} simulation that additionally respects
\begin{itemize}
    \item long-term collision \emph{paths} of pushing interactions,
    \item pulling hand interactions,
    \item and volumetric anatomical constraints.
\end{itemize}
Although the resulting $H^{\sia}_{T}$ appears more natural (\Sec{res:user}), our simulation \sia~is not real-time capable and, hence, potential applications are restricted. Therefore, our second objective is to train a neural network $\pre$ (\Sec{mtd:net}) that approximates $H^{\sia}_{T}$ while being real-time capable even on CPUs.

\subsection{Volumetric Template}\label{sec:mtd:temp}
In the remainder of this section, we will precisely state \sia~and \pre. However, as our approach is intended to reflect volumetric constraints, we first introduce an artist-created head template $\of{H, J, C, \tet{S}, \tet{J}, \tet{C}}$ as the foundation of \sia. The template includes the neutral head surface mesh $H \subset \tet{S}$ that encloses a soft tissue tetrahedral mesh $\tet{S}$. The two template surface meshes $J, C$ form the corresponding skull as jaw and cranium and are filled with respective tetrahedral meshes $\tet{J}$ and $\tet{C}$. All tetrahedral meshes are connected, and the surface vertices of $H$ can be addressed in $\tet{S}$ with the same indices.

\Fig{temp} b--d visualize all template components; all dimensions can be found in \App{dims}. The tessellation of $H$ is aligned with a full-body avatar  (\Fig{temp}a), which will be part of the code release to easily integrate NePHIM into other applications.

To register the volumetric template to a tracked person, we expect the neutral head surface $H$ of this person to be known. Then, we reposition the skull components by a dense linear model that we trained on the computed tomography dataset of Achenbach et al.~\shortcite{achenbach2018multilinear}. Formally, this model maps from the vertex positions of $H$ to the vertex positions of the jaw $J$ and the cranium $C$. The vertices of each tetrahedral mesh $\tet{S}, \tet{J}, \tet{C}$ are placed by radial basis function space warps ~\cite{botsch2005real} calculated \emph{from} the respective enclosing surfaces in the template \emph{to} the corresponding surfaces of the tracked person.

\subsection{Simulation}\label{sec:mtd:sim}
\algrenewcommand\algorithmicfunction{\textbf{Function}}
\begin{algorithm}[t]
\caption{Volumetric Physics-Based Simulation \sia}\label{alg:phy}
\begin{algorithmic}
\small
\Function{$\sia$}{$
     \sq{R}, 
    \sq{L},
    \sq{H},
    H$}

\LineComment{\Sec{mtd:temp} Register Template to Neutral }
\State \parbox[t]{\dimexpr\linewidth-\leftmargin-\labelsep-\labelwidth}{%A
 Register volumetric template (\Fig{temp}) to obtain the tracked person's volumetric head description $\of{H^{\reg}, J^{\reg}, C^{\reg}, \tet{S}^{\reg}, \tet{J}^{\reg}, \tet{C}^{\reg}}. $\strut}

\vspace{2mm}
\State $t = 1$
\While{$t \leq T$}\vspace{0mm}
\LineComment{\Sec{mtd:sim:ftt} Expression Fitting}
\State \parbox[t]{\dimexpr\linewidth-\leftmargin-\leftmargin-\labelsep-\labelwidth}{%
\textbf{Step 1} Determine the tracked tetrahedral meshes $\tet{S}^{\track}, \tet{J}^{\track}, \tet{C}^{\track}$ by aligning $\tet{S}^{\reg}, \tet{J}^{\reg}, \tet{C}^{\reg}$ with the tracked head surface $H^{\track}_{t}$ as described in \Eq{tracking}.\strut}

\vspace{1mm}\LineComment{\Sec{mtd:sim:fi} Detect Interactions}
\State \parbox[t]{\dimexpr\linewidth-\leftmargin-\leftmargin-\labelsep-\labelwidth}{%
\textbf{Step 2} Determine the push and pull target positions $C_\mathrm{push}, C_\mathrm{pull}$ as described in \Alg{push} and \Alg{pull}, respectively. \strut} 

\vspace{1mm}\LineComment{\Sec{mtd:sim:si} Simulate Interactions}
\State \parbox[t]{\dimexpr\linewidth-\leftmargin-\leftmargin-\labelsep-\labelwidth}{%
\textbf{Step 3} Determine the interaction tetrahedral meshes $\tet{S}^{\inter}, \tet{J}^{\inter}, \tet{C}^{\inter}$ by applying the push and pull targets $C_\mathrm{push}, C_\mathrm{pull}$ to $H^{\track}_{t}$ as described in \Eq{inter}.\strut}

\vspace{1mm}\LineComment{\Sec{mtd:sim:cor} Corrections}
\State \parbox[t]{\dimexpr\linewidth-\rightmargin-\leftmargin-\leftmargin-\labelsep-\labelwidth}{%
\textbf{Step 4} Determine the corrected tetrahedral meshes $\tet{S}^{\corr}, \tet{J}^{\corr}, \tet{C}^{\corr}$ by resolving remaining collisions $I_\mathrm{corr}$ as described in \Eq{correction}. Extract $H^{\sia}_{t}$ from $\tet{S}^{\corr}$.\strut}
\vspace{3mm}\State $t = t + 1$
\EndWhile

\LineComment{Return the simulated head surface}
\State \Return $H^{\sia}_{T}$
\EndFunction
\end{algorithmic}
\end{algorithm}

Building on the volumetric template, we can now continue with the detailed introduction of our physics-based simulation \sia. As \Alg{phy} outlines, \sia~conducts four steps that are described in separate subsections from \Sec{mtd:sim:ftt} to \Sec{mtd:sim:cor}. \Fig{method}b visualizes an exemplary cycle of all steps. Since we want to take long-term effects such as collision paths and skin pulling into account, it is not sufficient to consider only the last time step $T$ to determine $H^{\sia}_{T}$. Instead, we start at the beginning of the tracked sequences and run all four simulation steps consecutively for each time step $t$.

\subsubsection{Expression Fitting}\label{sec:mtd:sim:ftt}
As the initial simulation step, we deform the neutral volumetric tetrahedral meshes $\tet{S}^{\reg}, \tet{J}^{\reg},$ and $ \tet{C}^{\reg}$ in an anatomically plausible manner to fit the tracked surface $H^{\track}_{t}$ instead of the neutral surface $H$. To this end, we minimize a constraint-based energy in the projective dynamics simulation framework~\cite{bouaziz2014projective}. The first objective
\begin{equation}\label{eq:softd}
    E_\mathrm{target}\of{H^{\reg}, H^{\track}_{t}} = \norm{H^{\reg} - H^{\track}_{t}}^2
\end{equation}
attracts the surface vertices $H^{\reg}  \subset \tet{S}^{\reg}$ of the soft tissue tetrahedral mesh towards the tracked head surface. The second constraint
\begin{equation}
\begin{aligned}
    E_{\mathrm{strain}}\of{\mathbb{S}^{\reg}} = \!\sum_{\elm{t} \in \mathbb{S}^{\reg}}  & \!\min_{\mat{R} \in SO(3)} \left\Vert \nabla\of{\elm{t}} - \mat{R}\right\Vert^2_{F} \\
\end{aligned}
\end{equation}
models strain for each soft tissue tet $\elm{t} \in \mathbb{S}^{\reg}$. Here, $\mat{R} \in SO(3)$ denotes the optimal rotation, $\nabla\of{\elm{t}} \in \mathbb{R}^{3\times3}$ the deformation gradient of $\elm{t}$ (w.r.t. the neutral rest shape), and $\norm{\cdot}_F$ the Frobenius norm. Analogous to the soft tissue strain, we also add strain energies for the jaw $E_{\mathrm{strain}}\of{\mathbb{J}^{\reg}}$ and the cranium $E_{\mathrm{strain}}\of{\mathbb{C}^{\reg}}$. Overall, the weighted energy 
\begin{equation}\label{eq:tracking_energy}
    \begin{aligned}
    E_\func{tracked}\of{H^{\track}_{t}, \tet{S}^{\reg}, \tet{J}^{\reg}, \tet{C}^{\reg}} = \hspace{1mm}&w_\func{tar}E_\mathrm{target}\of{{H}^{\reg},
     H^{\track}_{t}} + \\
    & w_\mathbb{S}E_{\mathrm{strain}}\of{\mathbb{S}^{\reg}} + \\
    & w_\mathbb{J}E_{\mathrm{strain}}\of{\mathbb{J}^{\reg}} + \\ & w_\mathbb{C}E_{\mathrm{strain}}\of{\mathbb{C}^{\reg}} \\
    \end{aligned}
\end{equation}
is minimized. To reflect that both jaw and cranium are rigid, we set the weights $w_\mathbb{J}$ and $w_\mathbb{C}$ to a high value compared to $w_\mathrm{tar}$ and $w_\mathbb{S}$. The values of all weights and other simulation parameters can be found in \App{w}. The outputs of the optimization are the tracked tetrahedral meshes
\begin{equation}\label{eq:tracking}
    \of{\tet{S}^{\track}, \tet{J}^{\track}, \tet{C}^{\track}} = \underset{\tet{S}^{\reg}, \tet{J}^{\reg}, \tet{C}^{\reg}}{\mathrm{argmin}} \: E_\func{tracked}\of{H^{\track}_{t}, \tet{S}^{\reg}, \tet{J}^{\reg}, \tet{C}^{\reg}}.
\end{equation}

\subsubsection{Detect Interactions}\label{sec:mtd:sim:fi}
\begin{algorithm}[t]
\caption{Pushing Interaction}\label{alg:push}
\begin{algorithmic}
\small
\Function{\texttt{push}}{$
     H^{\sia}_{t - 1}, H^{\track}_{t}, 
    L^{\track}_{t-1}, L^{\track}_{t},
    R^{\track}_{t-1}, R^{\track}_{t}$}
\LineComment{Initialize linear movement directions}
\State $H_\mathrm{dir} = H^{\track}_{t} - H^{\sia}_{t-1} $
\State $L_\mathrm{dir} = L^{\track}_{t} - L^{\track}_{t-1} $
\State $R_\mathrm{dir} = R^{\track}_{t} - R^{\track}_{t-1} $
\LineComment{Initialize push targets}
\State $C_\mathrm{push} = \{\}$
\State $I = \{\}$
\vspace{1mm}
\LineComment{Iterate over linear movements}
\For{$\epsilon = 0; \epsilon \leq 1; \epsilon \mathrel{+}= $\text{\tiny${\Delta}$}\text{\normalsize$\epsilon$}}
    \LineComment{Iterate over head surface vertices}
    \For{$\elm{v}_i^H \in \of{H^{\sia}_{t - 1} + \epsilon \cdot H_\mathrm{dir}}$}
        \LineComment{Find collisions with left hand}
        \If{$\elm{v}_i^H$ collides with $\of{L^{\track}_{t-1} + \epsilon \cdot L_\mathrm{dir}}$ and $i \notin I$}
            \LineComment{Find nearest neighbor in current left hand}
            \State $\elm{v}_{\epsilon}^L = \texttt{nn}\of{\elm{v}_i^H, L^{\track}_{t-1} + \epsilon \cdot L_\mathrm{dir}}$
            \LineComment{Add same vertex of final left hand as target position}
            \State Add $\of{\elm{v}_{1}^L, i}$ to $C_\mathrm{push}$
            \State Add $i$ to $I$
        \EndIf
        \State \textbf{Repeat} the same \textbf{if}-clause for the right hand
    \EndFor 
\EndFor

\LineComment{Return the push targets}
\State \Return $C_\mathrm{push}$
\EndFunction
\end{algorithmic}
\end{algorithm}

\begin{algorithm}[t]
\caption{Pulling Interaction}\label{alg:pull}
\begin{algorithmic}
\small
\State \textit{Notation}
\State $\elm{c}^{L, f}_t$ \hspace{6.7mm} Cylinder of finger $f$ of the left hand $L$ at timestep $t$
\State $\texttt{len}$ \hspace{6.2mm} Length of a cylinder
\State $I_\mathrm{pull}[L, f]$ \hspace{1mm} Dictionary entry of key $L, f$, i.e., a list
\vspace{2mm}
\Function{\texttt{pull}}{$
     H^{\track}_{t} ,
    L^{\track}_{t-1}, L^{\track}_{t},
    R^{\track}_{t-1}, R^{\track}_{t}, I_\mathrm{pull}$}

\LineComment{Initialize pull targets}
\State $C_\mathrm{pull} = \{\}$

\LineComment{Check if new vertices are pulled per cylinder}
\For{$f = 1; f \leq 4; f \mathrel{+}= 1$}
    \LineComment{Pull only if cylinder gets smaller and is small enough}
    \If{$\texttt{len}\of{\elm{c}^{L, f}_t} < \texttt{len}\of{\elm{c}^{L, f}_{t-1}} \mathrm{and}~\texttt{len}\of{\elm{c}^{L, f}_t} < l_\mathrm{min}$}
    \LineComment{Check for each head vertex if inside cylinder}
    \For{$\elm{v}_i \in H^{\track}_{t}$}
        \If{$\elm{v}_i$ inside $\elm{c}^{L, f}_t$}
           \State Append $i$ to $I_\mathrm{pull}[L, f]$
        \EndIf
    \EndFor
    \EndIf
\EndFor

\LineComment{Check if vertices are no longer pulled per cylinder}
\For{$f = 1; f \leq 4; f \mathrel{+}= 1$}
    \If{$\texttt{len}\of{\elm{c}^{L, f}_t} \geq l_\mathrm{min}$}
        \State $I_\mathrm{pull}[L, f] = ()$
\EndIf
\EndFor
\LineComment{Calculate target positions of pulled vertices per cylinder by}\vspace{-1mm}
\LineComment{creating a $\texttt{ridge}$ per cylinder as defined in \App{ridge}}\vspace{-0mm}
\For{$f = 1; f \leq 4; f \mathrel{+}= 1$}
    \State Append $\texttt{ridge}\of{I_\mathrm{pull}[L, f], H^{\track}_{t}, \elm{c}^{L, f}_t} $ to $C_\mathrm{pull}$
\EndFor

\State \textbf{Repeat} same procedure for the right hand
\LineComment{Return the pull targets}
\State \Return $C_\mathrm{pull}$
\EndFunction
\end{algorithmic}
\end{algorithm}

\begin{figure}
    \centering
    \includegraphics[width=\linewidth]{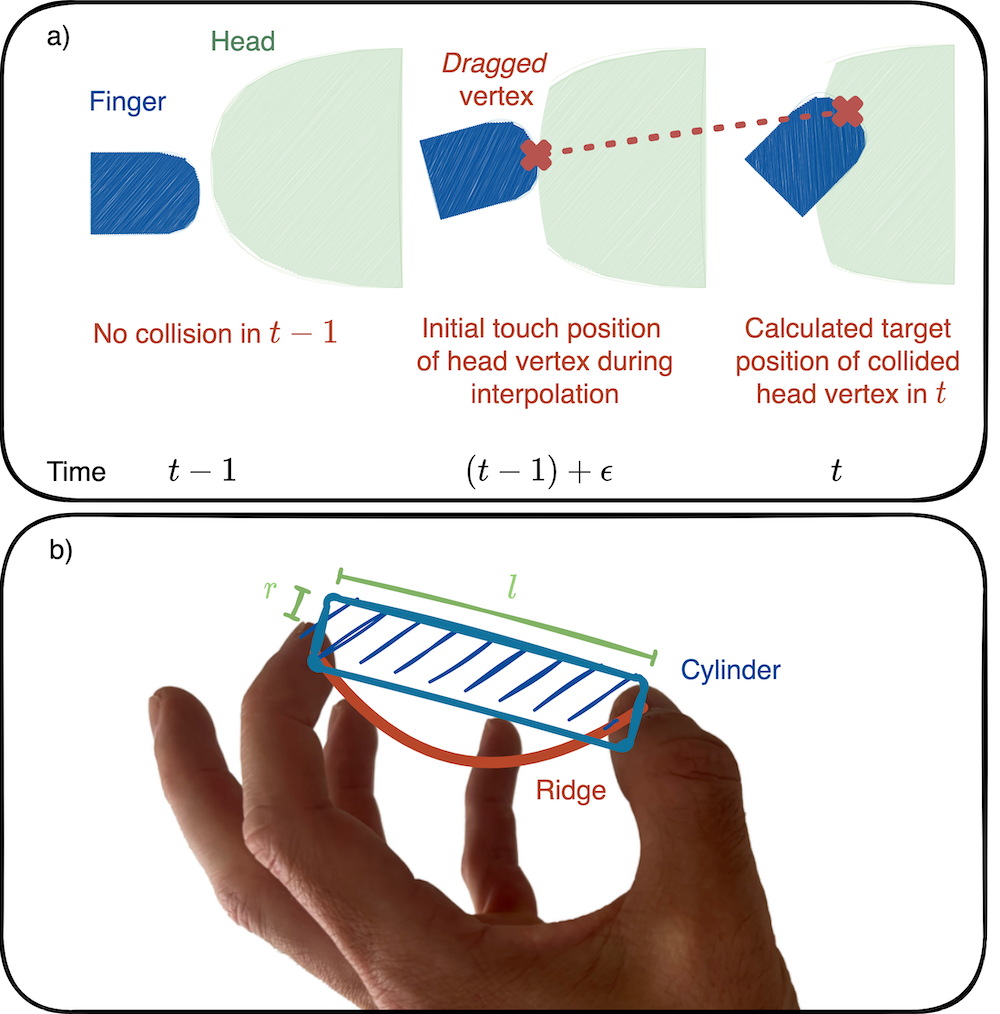}
        \caption{a) Visualization of \emph{pushing} as described in \Sec{mtd:sim:fi} and \Alg{push}. Here, $\epsilon$ is a substep between the time steps $t-1$ and $t$. b) Illustration of a finger cylinder with radius $r$,  length $l$, and an exemplary ridge shape that is used for \emph{pulling} as described in \Alg{pull}.}
    \label{fig:hand}
\end{figure}

Subsequently, we detect pushing and pulling head-hand interactions and translate them into target positions of the head vertices $C_\mathrm{push}$ and $C_\mathrm{pull}$, which we will simulate in the next step (\Sec{mtd:sim:si}). 

\noindent\textit{Pushing Interactions}
We first explain how we handle pushing in \sia. Roughly summarized, previous approaches \cite{shimada2023decaf,wu2024dice} would simply iterate over the vertices of the head surface $H^{\track}_{t}$ and if a vertex enters either the left hand $L^{\track}_{t}$ or the right hand $ R^{\track}_{t}$, it is moved in the direction of the corresponding inverted normal until the collision is resolved. Unfortunately, this strategy largely ignores temporal dependencies, and the normal direction only provides an imprecise collision resolution.

For this reason, we rely on the linear movements between $H^{\sia}_{t - 1}$ and $H^{\track}_{t}$, $L^{\track}_{t-1}$ and $L^{\track}_{t}$, as well as $R^{\track}_{t-1}$ and $R^{\track}_{t}$, i.e., between the previous simulated frame and the current tracked frame, as formally described in \Alg{push}. Expressed in words, we check at short intervals $\epsilon$ between the time steps $t-1$ and $t$ whether one of the two hands touches a vertex of the head. If so, the head vertex is dragged from the initial point of contact with the hand at $\epsilon$ to the same point on the hand at time $t$. Please see \Fig{hand}a for a visual explanation. Our way of resolving hand pushing is more \emph{natural} and incorporates long-term effects per construction. Although there are numerous more involved and time-consuming forms of continuous collision detection, these did not yield substantially better results in our experiments. 

\noindent\textit{Pulling Interactions} Pulling is considerably more challenging and has not been addressed in prior approaches. We present a heuristic in \Alg{pull} that does not require cumbersome friction calculations but, unfortunately, still has an elaborated notation. Yet, the foundational idea of our heuristics can easily be put into words. First, we form cylinders with radius $r$ between the fingertips of all fingers (index, middle, ring, little) and the thumb as illustrated in \Fig{hand}b. Then, for each cylinder, we determine whether it grabs, i.e., has shortened in length from time step $t-1$ to time step $t$. If so, and if the length falls below a minimum $l_\mathrm{min}$, all head vertices inside the cylinder at time $t$ are marked as \textit{pulled}. We maintain a dictionary $I_\mathrm{pull}$ over time that stores a list of the \textit{pulled} vertices for each finger. The target positions of the \textit{pulled} vertices $C_\mathrm{pull}$ are calculated so that they form smooth ridges within the cylinders (see \Fig{hand}b). The shape of the ridges imitates the skin's natural deformation due to pinching. A \textit{pulled} vertex is unmarked once the corresponding cylinder exceeds $l_\mathrm{min}$, i.e., the finger no longer grabs.

\subsubsection{Simulate Interactions}\label{sec:mtd:sim:si}
For applying the previously determined push and pull targets $C_\mathrm{push}$ and $ C_\mathrm{pull}$ to the tracked head $H^{\track}_{t}$, we again make use of a projective dynamics simulation on the fitted tetrahedral meshes $\tet{S}^{\track}, \tet{J}^{\track},$ and $ \tet{C}^{\track}$ (\Sec{mtd:sim:ftt}). Here, we establish anatomical plausibility similar as before by adding strain constraints $E_{\mathrm{strain}}\of{\tet{S}^{\track}}, E_{\mathrm{strain}}\of{\tet{J}^{\track}},$ and $ E_{\mathrm{strain}}\of{\tet{C}^{\track}}$ to the simulation. Also as before, we add \begin{equation}\label{eq:softd}
    E_\mathrm{target}\of{H^{\track}, H^{\track}_{t}} = \norm{H^{\track} - \of{H^{\track}_{t} + \frac{\alpha}{s} \of{H^{\sia}_{t - 1} - H^{\sia}_{t - 2}}}}^2
\end{equation}
to draw the surface vertices $H^{\track} \subset \tet{S}^{\track}$ of the soft tissue to the tracked surface. This time, however, including damped velocities of the head, where $s$ denotes the size of a time step. A low damping factor $\alpha$ adds natural-looking dynamic effects to the interactions.

New to the simulation are the target constraints
\begin{equation}
\begin{aligned}
    & E_\mathrm{push}\of{H^{\track}, C_\mathrm{push}} = \sum_{\of{\elm{p}, i} \in C_\mathrm{push}} \norm{\elm{p} - \elm{v}_i}^2, \\ & E_\mathrm{pull}\of{H^{\track}, C_\mathrm{pull}} = \sum_{\of{\elm{p}, i} \in C_\mathrm{pull}} \norm{\elm{p} - \elm{v}_i}^2, \\
\end{aligned}
\end{equation}
which draw interacting vertices $\elm{v}_i \in H^{\track}$ to their precalculated target position $\elm{p}$. Overall, the weighted energy 
\begin{equation}\label{eq:tracking_energy}
    \hspace{-2mm}\begin{aligned}
    E_\func{inter}\big(C_\mathrm{push}, C_\mathrm{pull},  & H^{\track}_{t}, \tet{S}^{\track},  \tet{J}^{\track}, \tet{C}^{\track}\big)  = \\   & 
    w_\func{push}E_\mathrm{push}\of{H^{\track}, C_\mathrm{push}} + \\
    &w_\func{pull}E_\mathrm{pull}\of{H^{\track}, C_\mathrm{pull}} + \\
    & w_\mathrm{tar}E_\mathrm{target}\of{{H}^{\track}, H^{\track}_{t}} + \\
    & w_\mathbb{S}E_{\mathrm{strain}}\of{\mathbb{S}^{\track}} + \\
    & w_\mathbb{J}E_{\mathrm{strain}}\of{\mathbb{J}^{\track}} + \\ &w_\mathbb{C}E_{\mathrm{strain}}\of{\mathbb{C}^{\track}} \\
    \end{aligned}
\end{equation}
is minimized, where we again set the weights $w_\mathbb{J}$ and $w_\mathbb{C}$ to a high value for approximating a rigid skull. Likewise, the weights $w_\func{push}$ and $w_\func{pull}$ are set to a high value to enforce the target positions, but lower as $w_\mathbb{J}, w_\mathbb{C}$. By balancing the previously mentioned weights, we achieve a more natural simulation since the bones do not bend in the case of tracking errors and too deeply penetrating hands. The outputs of the optimization are the interaction tetrahedral meshes
\begin{equation}\label{eq:inter}
\begin{aligned}
    \hspace{-2mm}\of{\tet{S}^{\inter}, \tet{J}^{\inter}, \tet{C}^{\inter}} = \!\!\underset{\tet{S}^{\track}, \tet{J}^{\track}, \tet{C}^{\track}} {\mathrm{argmin}}
    E_\func{inter}\big(& C_\mathrm{push}, C_\mathrm{pull},  \\ & \hspace{3mm}H^{\track}_{t}, \tet{S}^{\track}, \tet{J}^{\track}, \tet{C}^{\track} \big). \\
\end{aligned}
\end{equation}

\subsubsection{Corrections}\label{sec:mtd:sim:cor}
The preceding steps of \sia~do not fully resolve all head-hand collisions. For instance, the last step in \Sec{mtd:sim:si} allows soft tissue vertices that previously did not collide to move inside the hands. To correct most remaining colliding vertices, summarized with their indices in $I_\mathrm{corr}$, we perform the previous projective dynamics simulation again but add an additional constraint. This constraint
\begin{equation}
    E_\mathrm{corr}\of{\tet{S}^{\inter}, I_\mathrm{corr}} = \sum_{i \in I_\mathrm{corr}} \norm{\texttt{nn}\of{\elm{v}_i, L^{\track}_{t}, R^{\track}_{t}} - \elm{v}_i}
\end{equation}
draws each colliding vertex  $\elm{v}_i \in \tet{S}^{\inter}$ to the nearest neighbor $\texttt{nn}\of{\elm{v}_i, L^{\track}_{t}, R^{\track}_{t}}$ of $\elm{v}_i$ in the left or right hand $L^{\track}_{t}, R^{\track}_{t}$. The outputs of the optimization are the corrected tetrahedral meshes
\begin{equation}\label{eq:correction}
\hspace{-2mm}\begin{aligned}
    \of{\tet{S}^{\corr}, \tet{J}^{\corr}, \tet{C}^{\corr}} = &\underset{\tet{S}^{\inter}, \tet{J}^{\inter}, \tet{C}^{\inter}} {\mathrm{argmin}}  w_\mathrm{corr}E_\mathrm{corr}\of{\tet{S}^{\inter}, I_\mathrm{corr}} + \\
    & E_\func{inter}\of{C_\mathrm{push}, C_\mathrm{pull}, H^{\track}_{t}, \tet{S}^{\inter}, \tet{J}^{\inter}, \tet{C}^{\inter}}. \\
\end{aligned}
\end{equation}

The deformed surface $H^{\sia}_{t} = \sia\of{\mathcal{R}_t, 
\mathcal{L}_t,
\mathcal{H}_t,
H} \subset\tet{S}^{\corr} $ can now be extracted. After the four steps of \sia~described in Sections \ref{sec:mtd:sim:ftt}--\ref{sec:mtd:sim:cor} have been carried out consecutively for all time steps through to $T$, $H^{\sia}_T$ is obtained.

\subsection{Recursive Formulation}\label{sec:mtd:rec}
The previous description of \sia~serves the intuitive derivation, but suggests that the computational effort increases linearly with each additional frame. However, this is not the case, since by the design of \sia, we can rewrite \Eq{phy} recursively as 
\begin{equation}\label{eq:phy_rec}
\begin{aligned}
    {H}^{\sia}_{T} = \sia(
    & {L}^{\track}_{T}, {R}^{\track}_{T}, {H}^{\track}_{T}, \\
    & {L}^{\track}_{{T - 1}}, {R}^{\track}_{{T - 1}}, {H}^{\sia}_{{T - 1}}, \\
    & {H}^{\sia}_{{T - 2}}). \\
\end{aligned}
\end{equation}
Hence, we can reuse simulated frames instead of always simulating all time steps.
\subsection{Neural Simulation Approximation}\label{sec:mtd:net}
\begin{figure}
    \centering
    \includegraphics[width=1.0\linewidth]{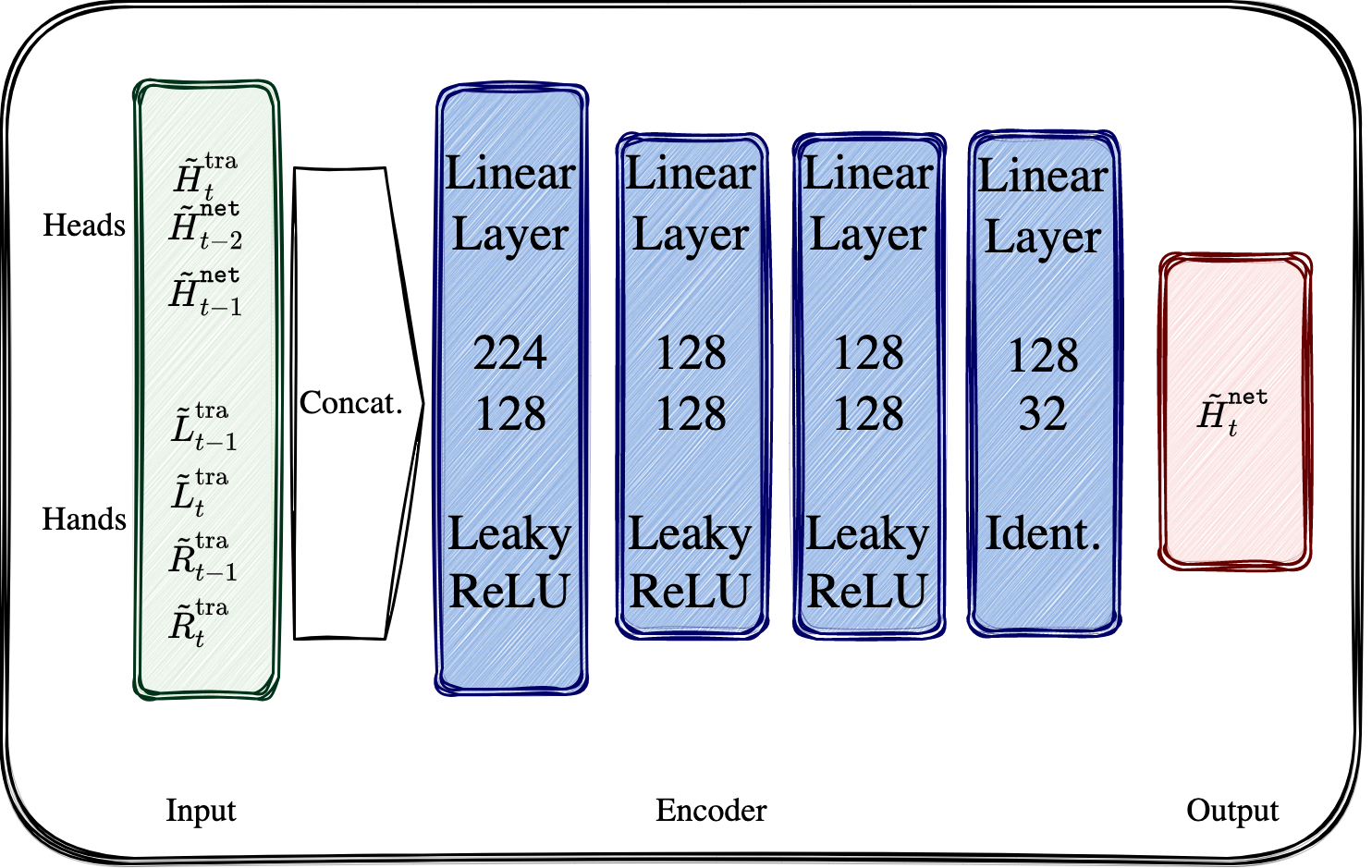}
    \caption{An overview of the efficient network architecture of \pre. Basically, a simple MLP with only 65.536 parameters.}
    \label{fig:network}
\end{figure}

As the derivations in the previous sections already indicate, \sia~is not real-time capable. Therefore, we construct \pre, a neural network that can be evaluated even on CPUs with 50 FPS (\Tab{runtimes}) and that closely approximates \sia. From the wide corpus of techniques that already exist for approximating physic-based simulations (\Sec{rw:lpy}), we adapt subspace neural physics (SNP) \cite{holden2019subspace} to our needs. Here, we only explain our adapted architecture of \pre, as the original publication extensively describes the training algorithm and we do not modify it.

The principle idea of SNP is to project all inputs and outputs into smaller linear subspaces (e.g. using principal component analysis (PCA)) and to train \pre~on the projection. In the following, the pedant of a variable in its respective subspace is referenced with an overlying tilde. The inputs of $\pre$ with regard to \sia~as defined in \Eq{phy_rec} are \begin{equation}
    L^{\track}_{T}, R^{\track}_{T}, H^{\track}_{T},
    L^{\track}_{{T - 1}}, R^{\track}_{{T - 1}}, H^{\sia}_{{T - 1}},
    H^{\sia}_{{T - 2}}.
\end{equation}
Consequently, we have to create PCA subspaces for the tracked left hand, the tracked right hand, the tracked head, and the simulated head, respectively. 
The overall training goal is then to minimize 
\begin{equation}
    \min_{\pre} \norm{\proj{H}^{\pre}_{T} - \proj{H}^{\sia}_{T}}^2,
\end{equation} where 
\begin{equation}
\begin{aligned}
    \proj{H}^{\pre}_{T} = \pre(
    & \proj{L}^{\track}_{T}, \proj{R}^{\track}_{T}, \proj{H}^{\track}_{T}, \\
    & \proj{L}^{\track}_{{T - 1}}, \proj{R}^{\track}_{{T - 1}}, \proj{H}^{\pre}_{{T - 1}}, \\
    & \proj{H}^{\pre}_{{T - 2}}). \\
\end{aligned}
\end{equation}
A visual illustration of the inputs and outputs of \pre~is depicted in \Fig{method}c and our architecture can be found in \Fig{network}. To recover ${H}^{\pre}_{T}$ from $\proj{H}^{\pre}_{T}$, the PCA of the simulated heads is applied. By selecting an appropriate number of components of the subspace, we prevent the loss of geometric details.

\section{Results}\label{sec::res}
The result section is organized as follows. First, we outline how we capture and process real head-hand interactions to form training and test data (\Sec{res:data}). The same subsection also contains a description of the resulting dataset and details on training and evaluation protocols. In \Sec{res:qual}, we discuss qualitative characteristics of the simulation $\sia$ and the approximation $\pre$~using visual examples. In \Sec{res:quan}, we examine quantitative characteristics and also take a closer look at running times as well as training times. Finally, we present the results of a user study (\Sec{res:user}) that supports the more natural perception of our approach.

\begin{figure*}
\captionsetup[subfigure]{labelformat=empty}
    \centering
     \begin{subfigure}[b]{0.45\linewidth}
             \centering
             \includegraphics[width=\linewidth]{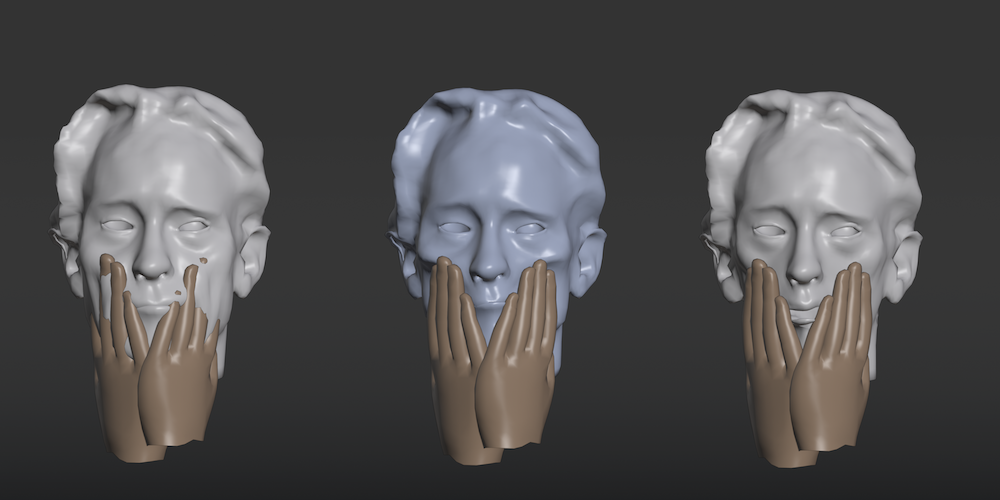}
     \end{subfigure}\hspace{0.07\textwidth}%
     \begin{subfigure}[b]{0.45\linewidth}
             \centering
             \includegraphics[width=\linewidth]{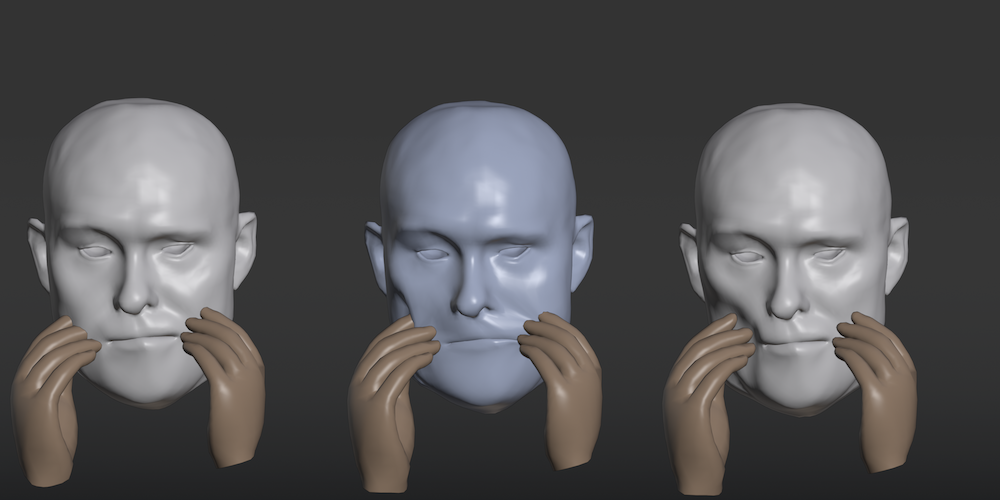}
      \end{subfigure}
% leave a blank line to change row         

     \begin{subfigure}[b]{0.45\linewidth}
             \centering
             \includegraphics[width=\linewidth]{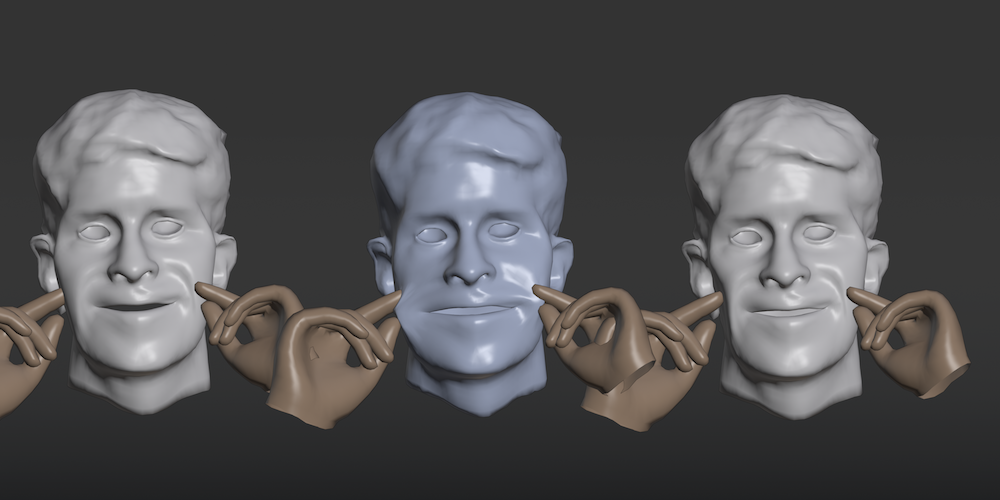}
     \end{subfigure}\hspace{0.07\textwidth}%
     \begin{subfigure}[b]{0.45\linewidth}
             \centering
             \includegraphics[width=\linewidth]{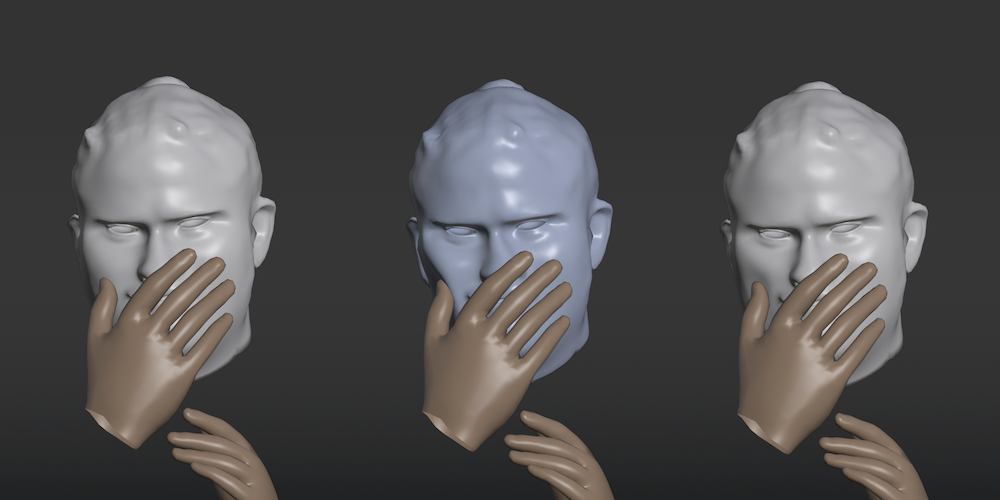}
      \end{subfigure}
% leave a blank line to change row         

     \begin{subfigure}[b]{0.45\linewidth}
             \centering
             \includegraphics[width=\linewidth]{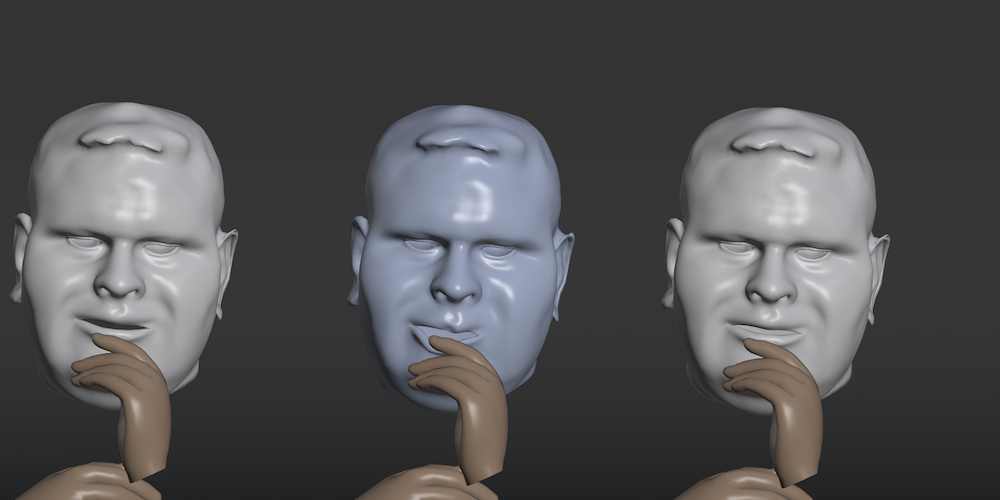}
             \caption{\raggedright \hspace{0.1cm} Tracked \hspace{2.2cm} Ours \hspace{1.5cm} Decaf \shortcite{shimada2023decaf}}
     \end{subfigure}\hspace{0.07\textwidth}%
     \begin{subfigure}[b]{0.45\linewidth}
             \centering
             \includegraphics[width=\linewidth]{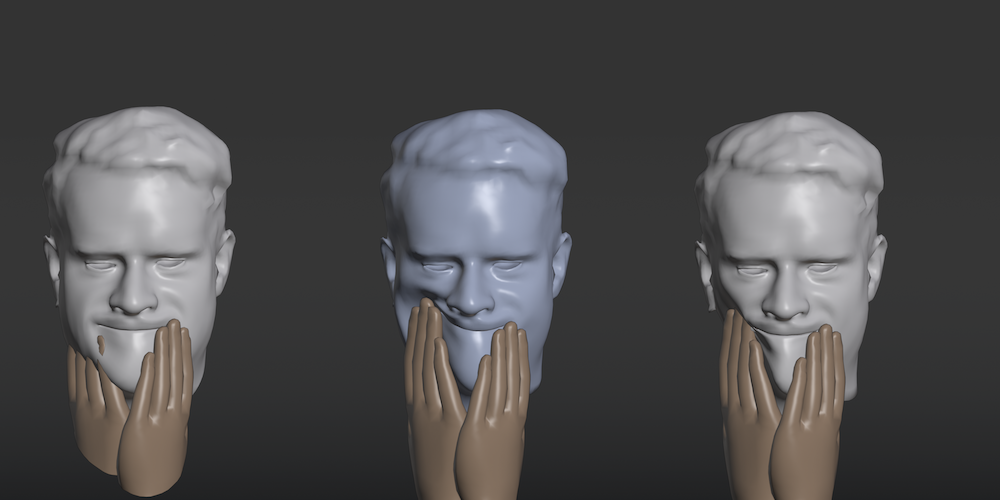}
             \caption{\raggedright \hspace{0.1cm} Tracked \hspace{2.2cm} Ours \hspace{1.5cm} Decaf \shortcite{shimada2023decaf}}
      \end{subfigure}
      \caption{The figure shows examples of our simulation $\sia$ and compares them to the tracked surfaces as well as the simulation of Decaf\cite{shimada2023decaf}. In the top left, for example, the advantage of simulating the skull becomes apparent near the cheekbone. In the top right image, a pulling interaction is shown and the lower images demonstrate the importance of collision paths.}
      \label{fig:qual}
\end{figure*}

\subsection{Dataset \& Training}\label{sec:res:data}
To capture real head-hand interactions, we use a multi--view rig consisting of 16 synchronized and calibrated XIMEA~\cite{xim} RGB cameras generating 12-megapixel images at 20 FPS.
In each captured image, we predict 2D landmarks for both hands and head using existing tracking methods~\cite{bulat2017far, zhang2020mediapipe}. For the hands, a landmark is predicted for each joint, each fingertip, and the wrists. For heads, we only capture the contours of the eyes and the mouth, as can be seen in \Fig{teaser}.
From the 2D landmarks, we generate 3D landmarks per frame using a basic bundle adjustment algorithm.

Since our simulation $\sia$~is conceptualized to work on tracked surfaces, the last step in the capturing pipeline is to fit appropriate template surfaces to the 3D landmarks.
Regarding the head, we initially create a high-resolution personalized head avatar for the recorded person with an automated 3D reconstruction and (nonlinear) template fitting pipeline. Afterward, we add linear blendshapes to the avatar by applying deformation transfer \cite{sumner2004deformation} to a set of template blendshapes. A professional digital artist designed the template blendshapes, which represent the 52 ARKit expressions~\cite{arkit}. Finally, we optimize per frame a set of corresponding blendshape weights, a translation vector, and a rotation matrix to fit the head surface to the respective 3D landmarks.
Regarding the hands, we adopt a similar approach. Here, however, we do not use a personalized hand model but optimize the pose and shape parameters of the MANO~\cite{MANO:SIGGRAPHASIA:2017} hand model to match the respective 3D landmarks. In contrast to the pose parameters, the shape parameters are the same for each frame. We use gradient descent as the optimizer for the surface fittings. \Fig{method}a illustrates all steps of the capturing pipeline.

The dataset we compiled contains up to 10 recordings of each of 8 individuals. The individuals are Caucasian males aged 26 to 54 with a body mass index ranging from slightly underweight to obese. Each recording lasts approximately 30 seconds and captures arbitrary head-hand interactions. In particular, we did not instruct the individuals on which hand movements or facial expressions they should perform. Overall, we captured, reconstructed, and simulated around 50.000 frames for this work. All of the following experiments concerning the neural network $\pre$~are always stated as an average of five runs, and we draw new train/test splits (90\%/10\%) for each run. All PCA subspaces have 32 components, which is sufficient in our case as we do not intend to generalize over large head or hand models.

\subsection{Qualitative Evaluation}\label{sec:res:qual}
\Fig{qual} displays instances of the simulation $\sia$ in comparison to the tracked surfaces as well as the simulation of the current state-of-the-art Decaf~\cite{shimada2023decaf}. Please note that we implemented the latter simulation ourselves as the announced implementations are not (yet) available. Decaf results sometimes appear slightly different to those from \cite{shimada2023decaf}, which mainly stems from the fact that our head avatars are more detailed than FLAME~\cite{li2017learning} and that we did not instruct the recorded persons which head-hand interactions they should perform. In the shown examples, it is especially striking that our temporal processing of hand pushes leads to effects such as a bent nose, a pushed-up mouth corner, or even a pushed-down lip. Moreover, the pulling of skin is readily recognizable and appears natural. None of these effects can be observed with the other methods. The accompanying video demonstrates the advantages of our method for dynamic scenes.

\begin{figure*}
\captionsetup[subfigure]{labelformat=empty}
    \centering
     \begin{subfigure}[b]{0.45\linewidth}
             \centering
             \includegraphics[width=\linewidth]{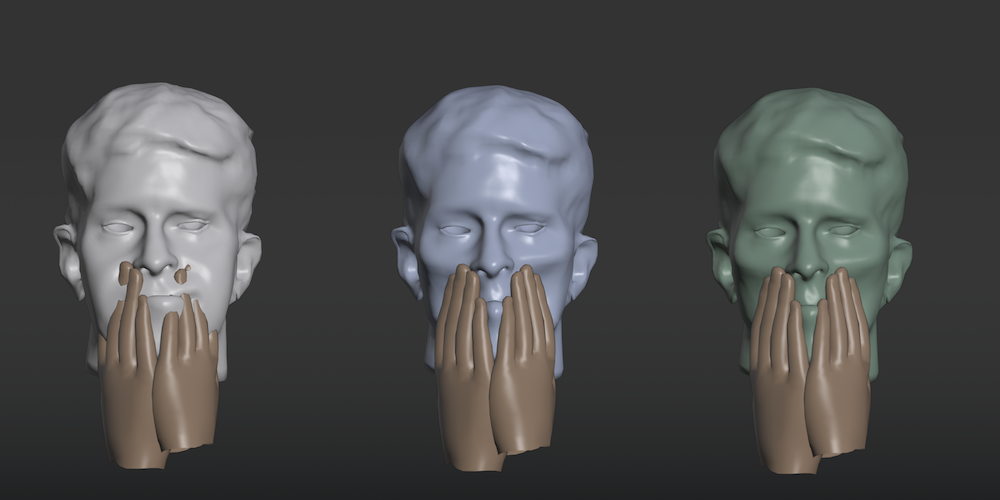}
     \end{subfigure}\hspace{0.07\textwidth}%
     \begin{subfigure}[b]{0.45\linewidth}
             \centering
             \includegraphics[width=\linewidth]{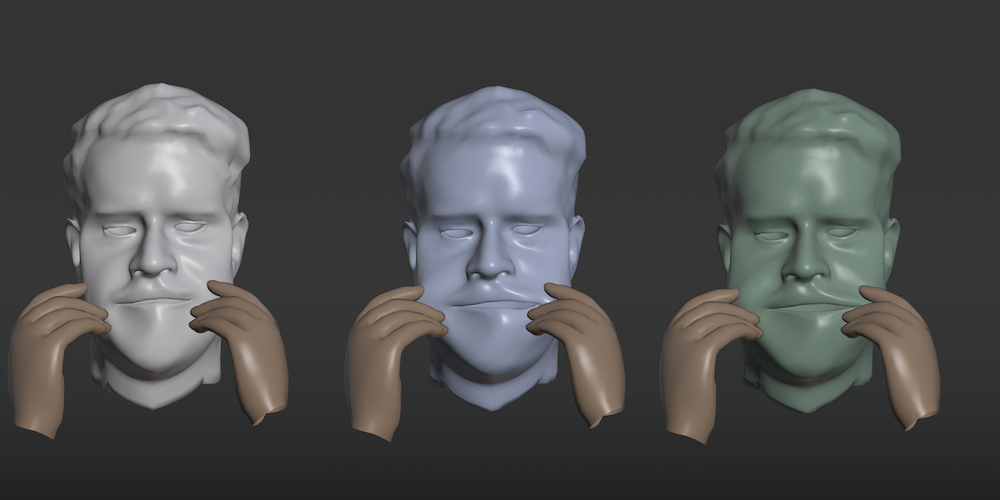}   
      \end{subfigure}
% leave a blank line to change row         

     \begin{subfigure}[b]{0.45\linewidth}
             \centering
             \includegraphics[width=\linewidth]{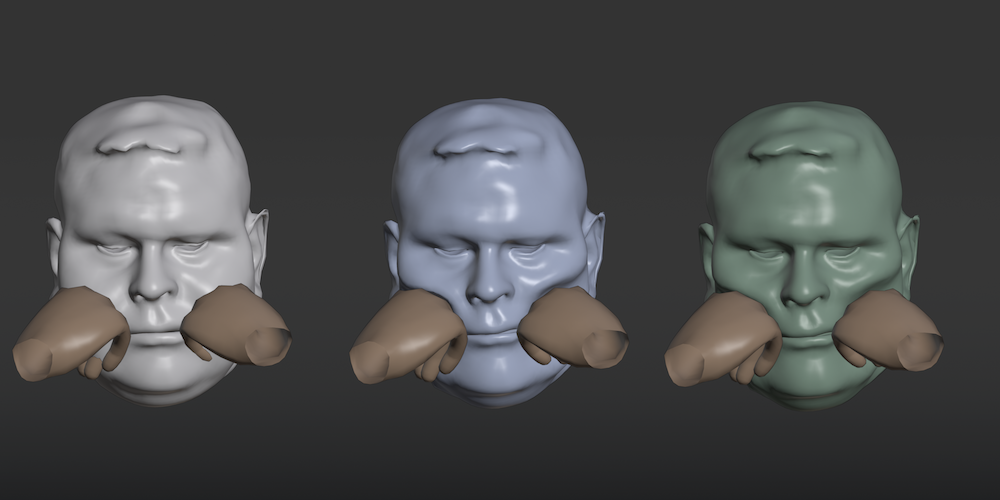}
             \caption{\raggedright \hspace{0.1cm} Tracked \hspace{2cm} Simulation \hspace{1.1cm} Approximation}  
     \end{subfigure}\hspace{0.07\textwidth}%
     \begin{subfigure}[b]{0.45\linewidth}
             \centering
             \includegraphics[width=\linewidth]{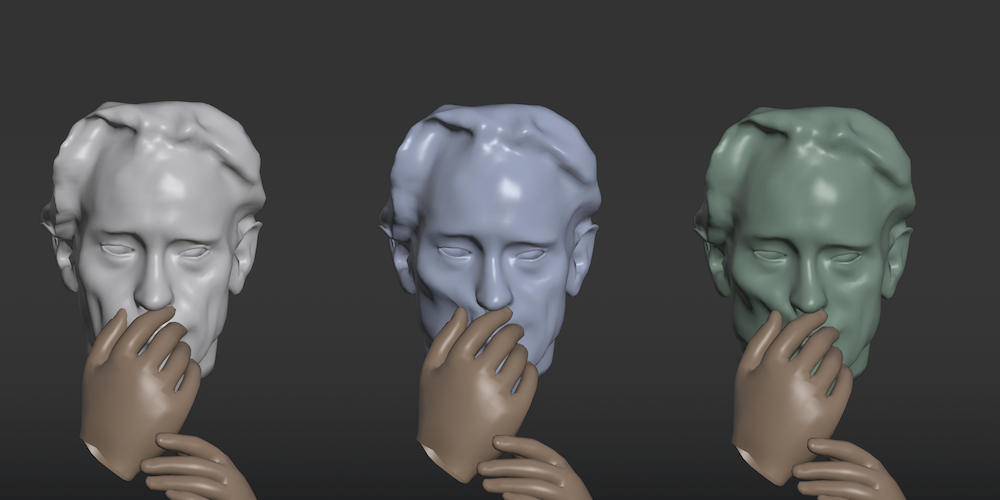}
             \caption{\raggedright \hspace{0.1cm} Tracked \hspace{2cm} Simulation \hspace{1.1cm} Approximation}
      \end{subfigure}
% leave a blank line to change row         

      \caption{The figure shows examples of our simulation $\sia$ along with the tracked surfaces as well as the learned neural approximation $\pre$ (trained on all identities in our dataset). As can be recognized, the quality of the approximation is independent of whether it is a pushing or a pulling interaction.}
      \label{fig:pred}
\end{figure*}
\Fig{pred} likewise shows examples of our simulation $\sia$~along with the tracked surfaces. This time, however, the learned approximation $\pre$ is depicted in comparison. For this purpose, we trained \pre~on all identities in our dataset simultaneously. Although minor discrepancies can be recognized, these do not appear to be decisive for visual perception. Moreover, the quality of the approximation is not affected by whether the head-hand interactions are pushing or pulling. As before, the accompanying video contains further examples.

\subsection{Quantitative Evaluation}\label{sec:res:quan}

\begin{table}[t]
\centering
\begin{tabular}{@{}ccccc@{}}
\toprule
\multirow{2}{*}{\# Identities} & \multicolumn{2}{c}{Subspace} & \multicolumn{2}{c}{Reconstruction} \\
                               & Train \textit{MSE}      & Test \textit{MSE}       & Train $\ell^2$          & Test $\ell^2$         \\ \midrule
One                            &       0.011       &    0.052               &                      0.02~\si{\centi\meter}       &           0.09~\si{\centi\meter}              \\
Eight                            &       0.041       &    0.056         &           0.04~\si{\centi\meter}       &           0.01~\si{\centi\meter}       \\ \bottomrule
\end{tabular}
\caption{Train and test errors of the neural approximation \pre~of the simulation \sia. The table is separated by the number of identities \pre~was trained on. The errors stated for \emph{one} identity are the average over separate networks for all identities in our dataset.}
\label{tab:error}
\end{table}
\begin{table}[t]
\centering
\begin{tabular}{@{}ccccc@{}}
\toprule
         \begin{tabular}[c]{@{}c@{}}\sia\\ CPU\end{tabular} & \begin{tabular}[c]{@{}c@{}}\pre\\ CPU\end{tabular} & \begin{tabular}[c]{@{}c@{}}\pre\\ GPU\end{tabular} &  \begin{tabular}[c]{@{}c@{}}Dataset\\ Creation\end{tabular} & \begin{tabular}[c]{@{}c@{}}Training\\ Time\end{tabular} \\
         \multicolumn{3}{c}{\textit{Per Frame}} &  \multicolumn{2}{c}{\textit{Eight / One Identities}} \\ \midrule
 876~\si{\milli\second}  & 19.2~\si{\milli\second}  & 5.1~\si{\milli\second} &  16~\si{\hour} /  2~\si{\hour}  &  4~\si{\hour} /  1.5~\si{\hour}                                                \\ \bottomrule
 
\end{tabular}
\caption{The average inference running times of the simulation \sia~and the neural approximation \pre~as well as the training time of \pre.}
\label{tab:runtimes}
\end{table}
This section mainly investigates the quantitative properties of the network  $\pre$. To begin with, we have a look at the approximation accuracy.
For this purpose, \Tab{error} summarizes average train and test subspace errors (mean squared error) of $\pre$~as well as the corresponding errors on the actual surfaces ($\ell^2$ error). There is also a breakdown by the number of identities with which we trained and tested $\pre$. The table indicates that the test errors are never greater than a millimeter, and our implementation of $\pre$~has sufficient capacity to generalize over several identities.

\Tab{runtimes} summarizes the average running and training times of $\sia$~and \pre. On the one hand, with a runtime of 876~\si{\milli\second} per frame on an AMD Ryzen Threadripper PRO 3995WX, $\sia$ is evidently not realtime-capable. On the other hand, $\pre$ can be executed not only on a consumer-grade GPU (NVIDIA RTX 3090) but also on a weaker CPU (Intel i5 12600K) with more than 50 FPS. In comparison, our implementation of the simulation of Decaf~\shortcite{shimada2023decaf} runs in 178~\si{\milli\second} per frame on the Threadripper CPU.  The entire pipeline, as shown in \Fig{method}, from data acquisition to training $\pre$~only takes about 20 hours for eight identities and 3.5 hours for one identity. We trained on four NVIDIA A6000 GPUs for four hours (eight identities) or one and a half hours (one identity). The short training time is mainly due to the efficient network architecture. 

\subsection{User Study}\label{sec:res:user}
\begin{figure}
    \centering
    \includegraphics[width=1.0\linewidth]{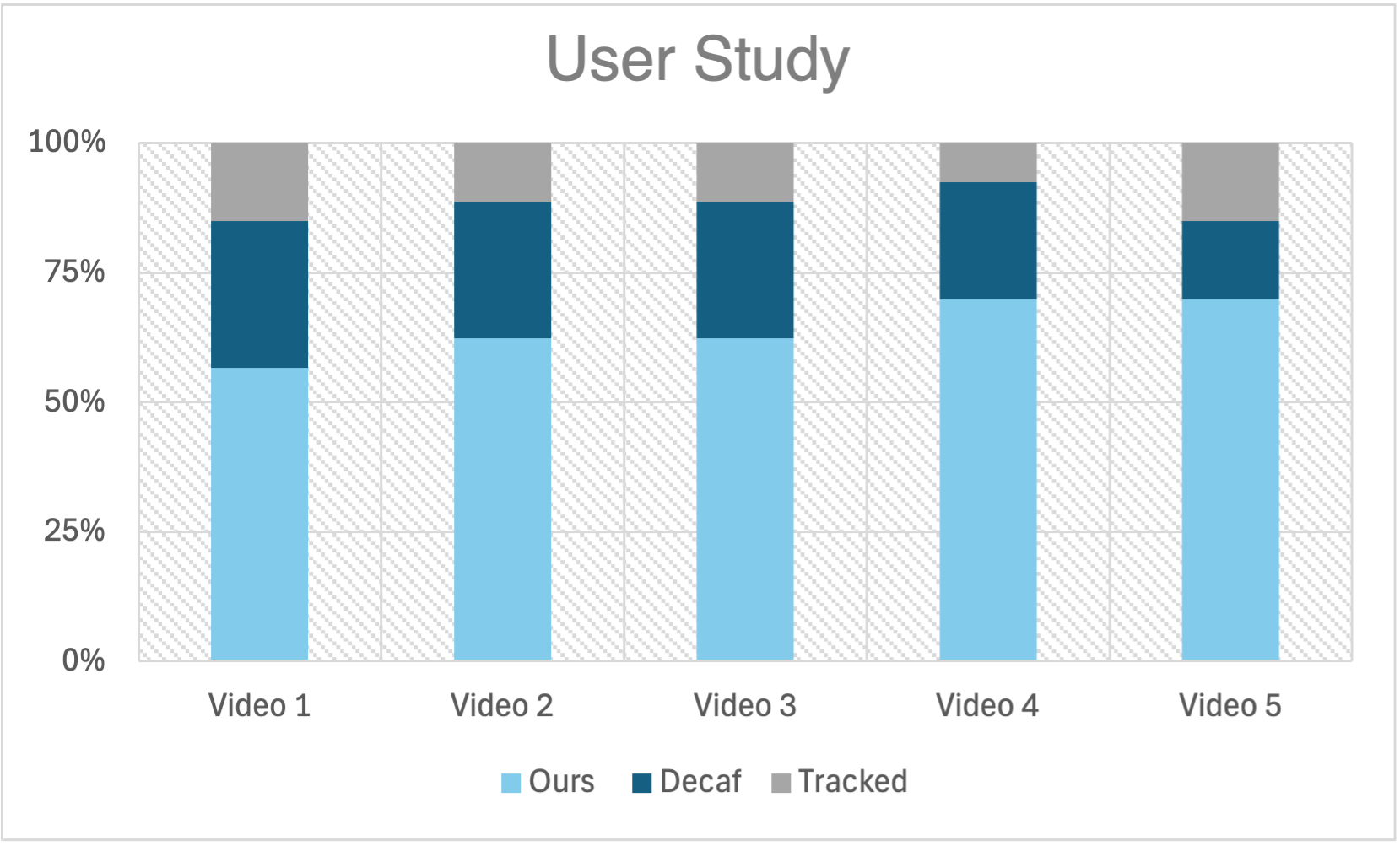}
    \caption{A user study among 53 participants supports that our approach is recognized as more natural. For each video shown in the user study, our approach received the most votes.}
    \label{fig:user}
\end{figure}

To support the qualitative results, we conducted an online user study with 53 participants from two universities. Each participant watched five random example videos that compared the tracked surfaces, the simulated surfaces of Shimada et al.~\shortcite{shimada2023decaf}, and our (neural) approximated surfaces as in \Fig{qual}. The videos are randomly drawn from the sequences in our dataset. Participants were asked to choose the most natural-looking of the three variants for each video. To avoid any bias, we rendered all surfaces in the same color and arranged the variants in random orders. To ensure independent documentation, we used \url{survio.com} for the technical implementation. 

\Fig{user} summarizes the outcome of the user study as the proportion of votes each variant received. Our approach achieved the most votes for all videos by a considerable margin.

\section{Limitations}\label{sec::lim}
\begin{figure}
    \centering
    \includegraphics[width=1.0\linewidth]{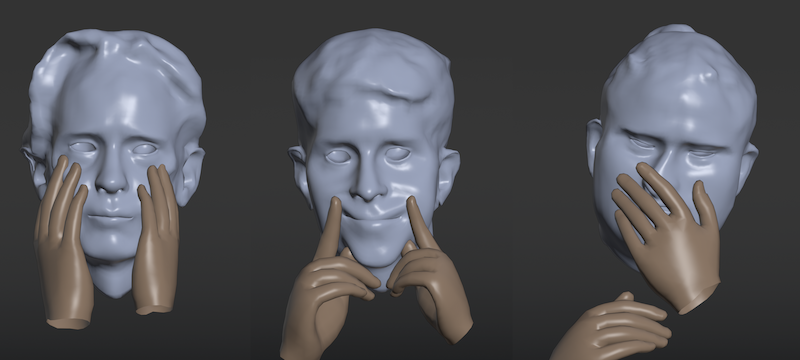}
    \vspace{-0.6cm}\caption*{\raggedright \hspace{1.0cm} (a) \hspace{2.4cm} (b) \hspace{2.4cm} (c)}\vspace{-0.2cm}
    \caption{(a) displays remaining collisions due to rigid bones, (b) self-intersections of lips, and (c) a too bendy nose due to missing cartilage.}
    \label{fig:limits}
\end{figure}

The most significant limitations of our approach result from missing details in the foundational physics-based simulation \sia~as demonstrated in \Fig{limits}. For instance, tracking errors can cause hands to move too deep into the head such that the skull is penetrated. In this case, we consider it more natural to not fully resolve collisions rather than bend bones (\Fig{limits}a). We also do not resolve self-collisions between lips or lips and teeth (\Fig{limits}b). Finally, in our anatomical head model, cartilage components are not sufficiently taken into account, causing the nose or ears to bend a bit too much when the hands push firmly (\Fig{limits}c).

Regarding the efficient approximation of \sia~by the neural network \pre, one can consider a lack of generalization over an extensive set of head shapes as a limitation. However, in contrast to previous work \cite{shimada2023decaf, wu2024dice} our focus is on personalized head avatars that exhibit a much higher level of detail and authenticity than commonly used head models ~\cite{li2017learning, feng2021learning}. Moreover, our experiments with multiple head shapes (\Sec{res:quan}) indicate generalization capacities of \pre~and the short training time of our approach should be sufficient in most scenarios to train $\pre$ to a given personalized head avatar.

Finally, a greater diversity in our dataset would be desirable. Although we cover a wide range of head shapes with different anatomical compositions, a more diverse coverage of genders and ethnicities would strengthen our results\footnote{We will increase the diversity and expand our dataset in the coming weeks.}.

\section{Conclusion}\label{sec::con}
In this work, we presented NePHIM, a neural physics-based head-hand interaction model. NePHIM extends previous interaction simulations \cite{shimada2023decaf, wu2024dice} with various features such as time-dependent collision paths, pulling of skin, and a higher anatomical precision. Comprehensive experiments and a user study show that our approach is perceived as being considerably closer to reality than the previous state-of-the-art~\cite{shimada2023decaf}. Furthermore, we successfully learned a neural approximator of the simulation that allows for rapid inference even on consumer-grade devices.

Nevertheless, we also discussed limitations that provide various starting points for future work. For instance, more detailed anatomical structures and physical properties may enhance the simulation. Moreover, learning the deformation of interactions directly from multi-view videos can contribute to further improvements.
\bibliographystyle{amsalpha-authoronly}  
\bibliography{bibo}

\appendix
\section{Template Dimensions}\label{sec:dims}
\begin{table}[H]
\small
\centering
\begin{tabular}{lcccccc}
\toprule
Mesh &  $H$  & $J$  & $C$ & $\tet{S}$ & $\tet{J}$ & $\tet{C}$\\
 \# Vertices  &  6688 &  886 &  4220 &  11001 & 899 & 3354 \\
 \# Faces / Tets & 13372 &  1768 & 8444 & 31456 &  4190 & 15634 \\
\bottomrule
\end{tabular}
\vspace{0mm}
\caption{The dimensions of all template components in our experiments.}
\label{tab:sizes}
\end{table}

\section{Weights \& Parameters}\label{sec:w}
\begin{table}[H]
\small
\centering
\begin{tabular}{ccccccc}
\toprule
$w_\mathrm{tar}$ & $ w_\tet{S}$ & $ w_\tet{J}$ & $ w_\tet{C}$ & $ w_\mathrm{push}$ & $ w_\mathrm{pull}$ &  $ w_\mathrm{corr}$  \\
$10^2$&$10^1$&$10^4$&$10^4$&$10^2$&$10^2$&$10^2$\\
\bottomrule
\end{tabular}
\vspace{0mm}
\caption{The weights of the physics-based simulations of \sia.}
\label{tab:weights}
\end{table}

\begin{table}[H]
\small
\centering
\begin{tabular}{cccccc}
\toprule
Proj. Dyn. Iterations & $\alpha$ & $l_\mathrm{min}$ & $r$ & $s$  & \text{\tiny${\Delta}$}\text{\tiny$\epsilon$}\\
10 & 0.01 & 2.5~\si{\centi\meter} & 0.5~\si{\centi\meter} & 50~\si{\milli\second} & $0.05$\\
\bottomrule
\end{tabular}
\vspace{0mm}
\caption{The parameters of the physics-based simulations of \sia.}
\label{tab:weights}
\end{table}

\section{Ridge Calculation}\label{sec:ridge}

\begin{algorithm}[H]
\caption{Cylinder Ridge}\label{alg:ridge}
\begin{algorithmic}
\small
\State \textit{Notation}
\State $\elm{c}$ \hspace{2mm} Cylinder
\State $H$ \hspace{2mm} Head Surface
\State $I$ \hspace{2mm} Indices of vertices that are in $\elm{c}$
\State $\elm{v}_i^H$ \hspace{2mm} Vertex $i$ of $H$
\State $\texttt{len}$ \hspace{2mm} Length of a cylinder
\State $\texttt{start}$ \hspace{2mm} Start of a cylinder
\State $\texttt{end}$ \hspace{2mm} End of a cylinder
\State $\texttt{plane}\of{\elm{r}, \elm{n}}$ \hspace{2mm} Plane in normal form
\State $\texttt{mean}$ \hspace{2mm} Mean of vertices
\State $\texttt{proj}\of{\elm{v}, \elm{p}}$ \hspace{2mm} Project vertex $\elm{v}$ on plane $\elm{p}$

\vspace{2mm}
\Function{\texttt{ridge}}{$
     I, H, \elm{c}$}

\LineComment{Initialize ridge targets}
\State $C_\mathrm{ridge} = \{\}$

\LineComment{Calculate mean of cylinder }
\State $\elm{r} = \of{\texttt{start}\of{\elm{c}} + \texttt{end}\of{\elm{c}}} / 2$
\LineComment{Calculate normal of cylinder plane}
\State $\elm{n} = \elm{r} - \texttt{mean}\of{H}$
\State $\elm{n} \mathrel{/}= \norm{\elm{n}}$
\LineComment{Calculate cylinder plane}
\State $\elm{p} = \texttt{plane}\of{\elm{r},  \elm{n}}$

\LineComment{Calculate targets}
\For{$i \in I$}
    \LineComment{Calculate plane position}
    \State $\elm{v}_i^\elm{p} = \texttt{proj}\of{\elm{v}_i^H, \elm{p}}$
    \LineComment{Calculate a height offset factor}
    \State $\kappa = \min\of{\norm{\texttt{start}\of{c} - \elm{v}_i^\elm{p}}, \norm{\texttt{end}\of{c} - \elm{v}_i^\elm{p}}} / \of{\texttt{len}\of{c} / 2}$
    \LineComment{Add ridge target}
    \State Add $\of{\elm{v}_{i}^p + \kappa \cdot \texttt{len}\of{c} \cdot \elm{n}, i}$ to $C_\mathrm{ridge}$
\EndFor

\LineComment{Return the ridge targets}
\State \Return $C_\mathrm{ridge}$
\EndFunction
\end{algorithmic}
\end{algorithm}
\end{document}